\documentclass[twocolumn,showpacs,amsmath,amssymb]{revtex4}
\usepackage{graphics}
\usepackage{epsfig}
\usepackage{amsmath}
\usepackage{gensymb}

\begin{document}

\title{Search for Lorentz and CPT Violation Effects in Muon Spin
 Precession}
 
\author{
G.W.~Bennett$^{2}$, B.~Bousquet$^{10}$, H.N.~Brown$^2$, G.~Bunce$^2$,
R.M.~Carey$^1$, P.~Cushman$^{10}$, G.T.~Danby$^2$, P.T.~Debevec$^8$,
M.~Deile$^{13}$, H.~Deng$^{13}$, W.~Deninger$^8$, S.K.~Dhawan$^{13}$,
V.P.~Druzhinin$^3$, L.~Duong$^{10}$, E.~Efstathiadis$^1$, F.J.M.~Farley$^{13}$,
G.V.~Fedotovich$^3$, S.~Giron$^{10}$, F.E.~Gray$^8$, D.~Grigoriev$^3$,
M.~Grosse-Perdekamp$^{13}$, A.~Grossmann$^7$, M.F.~Hare$^1$, D.W.~Hertzog$^8$,
X.~Huang$^1$, V.W.~Hughes$^{13,*}$, M.~Iwasaki$^{12}$,
K.~Jungmann$^{6,7}$, D.~Kawall$^{13}$, M.~Kawamura$^{12}$, B.I.~Khazin$^3$,
J.~Kindem$^{10}$, F.~Krienen$^1$, I.~Kronkvist$^{10}$, A.~Lam$^1$,
R.~Larsen$^2$, Y.Y.~Lee$^2$, I.~Logashenko$^{1,3}$, R.~McNabb$^{10,8}$,
W.~Meng$^2$, J.~Mi$^2$, J.P.~Miller$^1$, Y. Mizumachi$^{9,11}$, W.M.~Morse$^2$,
D.~Nikas$^2$, C.J.G.~Onderwater$^{8,6}$, Y.~Orlov$^4$, C.S.~\"{O}zben$^{2,8}$,
J.M.~Paley$^1$, Q.~Peng$^1$, C.C.~Polly$^8$, J.~Pretz$^{13}$, R.~Prigl$^{2}$,
G.~zu~Putlitz$^7$, T.~Qian$^{10}$, S.I.~Redin$^{3,13}$, O.~Rind$^1$,
B.L.~Roberts$^1$, N.~Ryskulov$^3$, S.~Sedykh$^8$, Y.K.~Semertzidis$^2$,
P.~Shagin$^{10}$, Yu.M.~Shatunov$^3$, E.P.~Sichtermann$^{13}$, E.~Solodov$^3$,
M.~Sossong$^8$, A.~Steinmetz$^{13}$, L.R.~Sulak$^{1}$, C.~Timmermans$^{10}$,
A.~Trofimov$^1$, D.~Urner$^8$, P.~von~Walter$^7$, D.~Warburton$^2$,
D.~Winn$^5$, A.~Yamamoto$^9$ and
D.~Zimmerman$^{10}$ \\
(Muon $(g-2)$ Collaboration) }

\affiliation{
\mbox{$\,^1$Department of Physics, Boston University, Boston, MA 02215}\\
\mbox{$\,^2$Brookhaven National Laboratory, Upton, NY 11973}\\
\mbox{$\,^3$Budker Institute of Nuclear Physics, 630090 Novosibirsk, Russia}\\
\mbox{$\,^4$Newman Laboratory, Cornell University, Ithaca, NY 14853}\\
\mbox{$\,^5$Fairfield University, Fairfield, CT 06430}\\
\mbox{$\,^6$ Kernfysisch Versneller Instituut, University of Groningen,} \\
\mbox{NL-9747 AA, Groningen, The Netherlands}\\
\mbox{$\,^7$ Physikalisches Institut der Universit\"at Heidelberg, 69120 Heidelberg, Germany}\\
\mbox{$\,^8$ Department of Physics, University of Illinois at Urbana-Champaign, Urbana, IL 61801}\\
\mbox{$\,^9$ KEK, High Energy Accelerator Research Organization, Tsukuba, Ibaraki 305-0801, Japan}\\
\mbox{$\,^{10}$Department of Physics, University of Minnesota,
Minneapolis, MN 55455}\\
\mbox{$\,^{11}$ Science University of Tokyo, Tokyo, 153-8902, Japan}\\
\mbox{$\,^{12}$ Tokyo Institute of Technology, 2-12-1 Ookayama, Meguro-ku, Tokyo, 152-8551, Japan}\\
\mbox{$\,^{13}$ Department of Physics, Yale University, New Haven, CT 06520} }

\date{\today}

\begin{abstract}
The spin precession frequency of muons stored in the $(g-2)$ 
storage ring has been analyzed for evidence of Lorentz and CPT violation.
Two  Lorentz and CPT violation signatures were searched for:
a nonzero $\Delta\omega_{a}$ (=$\omega_{a}^{\mu^{+}}-\omega_{a}^{\mu^{-}}$);
 and a sidereal variation of $\omega_{a}^{\mu^{\pm}}$. No significant
 effect is found, and the following limits on the 
standard-model extension parameters are obtained:
$b_{Z} =-(1.0 \pm 1.1)\times 10^{-23}$~GeV;
$(m_{\mu}d_{Z0}+H_{XY}) = ( 1.8 \pm 6.0 \times 10^{-23})$~GeV; and
the 95\% confidence level limits
$\check{b}_{\perp}^{\mu^{+}}< 1.4 \times 10^{-24}$~GeV and
$\check{b}_{\perp}^{\mu^{-}} < 2.6 \times 10^{-24}$~GeV.
\end{abstract}

\maketitle

The minimal standard model of particle physics is  Lorentz and
 CPT invariant. Since the standard model is expected to
 be the low-energy limit of
a more fundamental theory such as string theory
 that incorporates gravity, Lorentz and CPT invariance might
 be broken spontaneously
 in the underlying theory~\cite{Kostel89}.
At low energies, the Lorentz and CPT violation signals are expected to be
small but perhaps observable in precision experiments.

To describe the effects of spontaneous breaking of
Lorentz and CPT invariance,  Colladay and 
Kosteleck\'{y}~\cite{Colladay97-98} 
proposed a general standard model extension  that can be viewed
as the low-energy limit of a Lorentz covariant theory. Lorentz and CPT
violating terms are introduced into the Lagrangian as a way of modeling
the effect of spontaneous symmetry breaking in the underlying fundamental
theory.  Other conventional
properties of quantum field theory such as gauge invariance,
renormalizability and energy conservation are maintained, and
the effective theory can 
 be quantized by the conventional approach. In a subsequent paper,
Bluhm, Kosteleck\'{y} and Lane discussed 
specific precision experiments with muons that could be sensitive to the
CPT and Lorentz violating interactions~\cite{Bluhm00}.

In this letter we present our analysis for 
CPT and Lorentz violating interactions in the anomalous spin 
precession frequency, $\omega_a$, 
of the muon moving in a magnetic field.  In experiment 
E821~\cite{Bennett06} at the Brookhaven
National Laboratory Alternating Gradient Synchrotron, 
muons are stored in a  magnetic storage ring that uses 
electrostatic quadupoles for vertical focusing. The storage ring 
has a highly uniform magnetic field with a central value of
 $B_0 = 1.45$~T, and a central radius
of $\rho = 7.112$~m. Polarized
muons are injected into the storage ring, and the positrons (electrons) 
from the parity-violating decay $\mu^{+(-)} \rightarrow e^{+(-)}\ \bar \nu_\mu 
( \nu_\mu)\  \nu_e (  \bar \nu_e) $ carry average
information on the muon spin
direction at the time of the decay.  Twenty-four electromagnetic calorimeters 
around the ring provide the arrival time and energy of the decay positrons.
As the muon spin precesses relative to the momentum with the frequency
$\omega_a$,  which is the the difference between the
spin precession frequency $\omega_S$ and the momentum (cyclotron) frequency
$\omega_C$~\cite{Bennett06}, the number of 
high-energy positrons is modulated by
$\omega_a$.  
In the approximation that $\vec \beta \cdot \vec B=0$,
\begin{equation}
\vec{\omega}_{a} =-\frac{q}{m}\left[a_{\mu}\vec{B}
-\left(
a_{\mu}-\frac{1}{\gamma^{2}-1}
\right)
\vec{\beta}\times\vec{E}\right] \, .
\label{eq:omegaa}
\end{equation}
where  $\gamma=1/\sqrt{1- \beta^{2}}$. The anomaly
$a_{\mu}$ is related to the spin $g$-factor
by $a_{\mu} = (g_\mu -2)/2$, with the magnetic moment given by 
$\vec \mu = g (q/ 2m) \vec s$, and $q = \pm e$.
The `magic' momentum of $p_m=3.09$~GeV/c, 
($\gamma_m=29.3$), was used in E821 so that the second term in 
Eq.~\ref{eq:omegaa} vanishes, and the electric field
does not contribute to $\omega_a$.

The magnetic field is measured using
nuclear magnetic resonance (NMR) techniques~\cite{Bennett06}.  The 
NMR frequency,  averaged over both the muon distribution and the
data collection period, is tied 
through calibration to the Larmor frequency for a
{\it free} proton, and is denoted by $\omega_p$~\cite{Bennett06}.  Thus
two frequencies are measured, $\omega_a$ and $\omega_p$.

In the analysis presented here,
the muon frequency $\omega_{a}$ is obtained from a fit of the 
positron (electron) arrival-time spectrum $N(t)$ to the 5-parameter function
\begin{equation}
N(t)=
N_{0}e^{-\frac{t}{\gamma\tau}}\left( 1+A\cos(\omega_{a}t+\phi) \right) \, .
\label{eq:fivepar}
\end{equation}  
 The normalization $N_{0}$, asymmetry $A$ and phase $\phi$ depend on the
 chosen energy threshold E. While more complicated fitting functions are
used in the analysis for the muon anomaly~\cite{Bennett06}, they represent
small deviations from Eq. \ref{eq:fivepar} and are not
 necessary for the CPT analysis.
The anomalous magnetic moment $a_{\mu}$ is
 calculated from $a_{\mu}={\mathcal R} /  (\lambda- {\mathcal R})$,
where ${\mathcal R} \equiv \omega_{a}$/$\omega_{p}$
and
$\lambda \equiv \mu_{\mu}/\mu_{p}=3.183\ 345\ 39(10)$  ($\pm 30$ ppb)
is the muon-proton magnetic moment ratio~\cite{lambda}.
 
For the muon, the Lorentz
 and CPT violating  terms in the Lagrangian are~\cite{Bluhm00}
\begin{equation}
\begin{split}
L^{\prime}=&-a_{\kappa}\bar{\psi}\gamma^{\kappa}\psi 
-b_{\kappa}\bar{\psi}\gamma_{5}\gamma^{\kappa}\psi-\frac{1}{2}
H_{\kappa\lambda}\bar{\psi}\sigma^{\kappa\lambda}\psi \\
&+\frac{1}{2}ic_{\kappa\lambda}\bar{\psi}\gamma^{\kappa}
 \stackrel{\textstyle\leftrightarrow}{D^{\lambda}} \psi 
+\frac{1}{2}id_{\kappa\lambda}\bar{\psi}\gamma_{5}\gamma^{\kappa}
 \stackrel{\textstyle\leftrightarrow}{D^{\lambda}} \psi,
\end{split}
\end{equation}
where $iD_{\lambda}\equiv i\partial_{\lambda}-qA_{\lambda}$,
and the small parameters $a_{\kappa}$,
 $b_{\kappa}$, $H_{\kappa\lambda}$, $c_{\kappa\lambda}$  
$d_{\kappa\lambda}$ represent the Lorentz and CPT
 violation.  All terms violate Lorentz invariance with 
$-a_{\kappa}\bar{\psi}\gamma^{\kappa}\psi$ and
$-b_{\kappa}\bar{\psi}\gamma_{5}\gamma^{\kappa}\psi$
CPT odd; all the other terms are CPT even.
In this model the conventional figure of merit
 $r_{g}^{\mu}\equiv|g_{\mu^{+}}-g_{\mu^{-}}|/g_{average}$ is 
zero at leading order; however  effects on the anomalous
 spin precession frequency $\omega_a$ do exist in
lowest-order~\cite{Bluhm00}. 
The frequency $\omega_a$ is proportional to the magnetic field 
and therefore to $\omega_p$, so
the sidereal variation of ${\mathcal R}=\omega_a/ \omega_p$ is
analyzed, rather than $\omega_a$ directly.

To compare results from different experiments, it is convenient to 
work in the non-rotating  standard celestial equatorial
 frame $\{\hat{X}, \hat{Y}, \hat{Z}\}$~\cite{Kostel99}. 
The $\hat{Z}$ axis is along the earth's rotational north pole,
with the $\hat{X}$ 
and $\hat{Y}$ axes lying in the plane of the earth's equator. The 
precession of the earth's rotational axis can be ignored because 
its precession period is 26,000 years. In this frame the correction 
to the (standard model) muon anomalous precession 
frequency $\omega_{a}^{\mu^{\pm}}$ in Eq.~\ref{eq:omegaa}
is calculated to be
\begin{equation} 
\delta\omega_{a}^{\mu^{\pm}} \approx
 2\check{b}_{Z}^{\mu^{\pm}}\cos\chi+2(\check{b}_{X}^{\mu^{\pm}}\cos\Omega t +
 \check{b}_{Y}^{\mu^{\pm}}\sin\Omega t)\sin\chi \, ;
\label{eq:deltaomega}
\end{equation}
\begin{equation}
\check{b}_{J}^{\mu^{\pm}}\equiv
\pm\frac{b_{J}}{\gamma}+m_{\mu}d_{J0}+\frac{1}{2}\varepsilon_{JKL}H_{KL}.
\mbox{ (J=X, Y, Z)}\, ,
\label{eq:bpm}
\end{equation}
where $\chi$ is the geographic colatitude $(=90^{\circ}\mbox{-
 latitude})$
 of the experiment location.
For E821, $\chi_{\tiny BNL} =49.1^\circ $.
The sidereal angular frequency is $\Omega=2\pi/T_{s}$, where
 $T_{s} \approx \mbox{ 23 hours 56 minutes}$.
 Eq.~\ref{eq:deltaomega} predicts two signatures of Lorentz
 and CPT violation: a difference between the time averages of
 $\omega_a^{\mu^{+}}$ and $\omega_a^{\mu^{-}}$, 
and oscillations in the values of $\omega_a^{\mu^+}$ and 
$\omega_a^{\mu^-}$ at the sidereal angular frequency.  

The E821 muon $(g-2)$ data have been analyzed for 
 these two signatures. Data from
 the 1999 run ($\mu^{+}$)~\cite{Brown01}, 2000 ($\mu^{+}$)~\cite{Bennett02}
 and 2001 ($\mu^{-}$)~\cite{Bennett04} were used. 
  Since bounds on clock comparisons
of $^{199}$Hg and $^{133}$Cs~\cite{Kostel99,Berglund95} place limits
on the Lorentz-violating energy shifts in the proton precession frequency
($\omega_p$) 
of $\sim 10^{-27}$~GeV, any shifts in the NMR measurements are at
the mHz level, which is negligible compared to the uncertainty in 
the amplitude of any sidereal variation in $\omega_a$.
 A feedback system based on the reading from NMR probes at 
fixed locations around the ring keeps the field constant to within 3 ppm. 
Both the proton frequency $\omega_{p}$ and the muon frequency
$\omega_a$ are measured
relative to a clock  stabilized to a LORAN C 10 MHz
 frequency standard\cite{LORAN}, 
via a radio signal. The LORAN C frequency standard is
 based on Cesium hyperfine transitions with $m_{F}=0$, which are insensitive
 to orientation of the clock. The Michelson-Morley type 
experiment of Brillet and Hall~\cite{Brillet79}
 establishes that the fractional frequency shift, $\Delta f/f$ of the
LORAN radio signals due to the earth's rotation
is less than $10^{-14}$, so at the level
of the precision of this experiment, the measured value of
$\omega_p$ is independent of sidereal time.  Moreover, since
LORAN C is the frequency standard for $\omega_a$ as well, any 
sidereal variation in the LORAN C standard would cancel in $\mathcal R$.

Comparison of the time averages of $\omega_{a}$ over many sidereal days
(Eq.~\ref{eq:deltaomega}) gives
$\Delta\omega_{a}\equiv
\langle\omega_{a}^{\mu^{+}}\rangle-\langle\omega_{a}^{\mu^{-}}\rangle
= ({4b_{Z}}/{\gamma})\cos\chi$.
For measurements made at different $\chi$ and/or $\omega_p$,
\begin{equation}
\begin{split}
\Delta {\mathcal R} = &\frac{2b_{Z}}{\gamma}
\left( \frac{\cos\chi_{1}}{\omega_{p1}}+
\frac{\cos\chi_{2}}{\omega_{p2}}\right)\\
&+2(m_{\mu}
d_{Z0}+H_{XY})
\left(\frac{\cos\chi_{1}}{\omega_{p1}}-\frac{\cos\chi_{2}}{\omega_{p2}}\right).
\label{eq:twochi}
\end{split}
\end{equation}
The colatitudes for the E821 $\mu^+$ and $\mu^-$  measurements were
identical, and the slightly different values of
$\omega_p$~\cite{Bennett06} can be neglected.
The E821 result~\cite{Bennett06} is
 ${\Delta {\mathcal R}}= -(3.6\pm3.7)\times 10^{-9}$, which 
corresponds to  
\begin{equation}
\begin{split}
b_{Z}&= -(1.0\pm1.1) \times 10^{-23}~{\rm GeV}\,\ {\rm or} \\
r_{\Delta\omega_{a}}&\equiv \frac{\Delta\omega_{a}}{m_{\mu}} 
=-(8.7 \pm 8.9)\times 10^{-24}
\, ,
\end{split}
\label{eq:821test1}
\end{equation}
a factor of 22 improvement over the limit that can be
obtained from the CERN muon $(g-2)$ experiment~\cite{Bailey79}.

\begin{figure}[]
\begin{center}
{\includegraphics[width=0.5\textwidth]{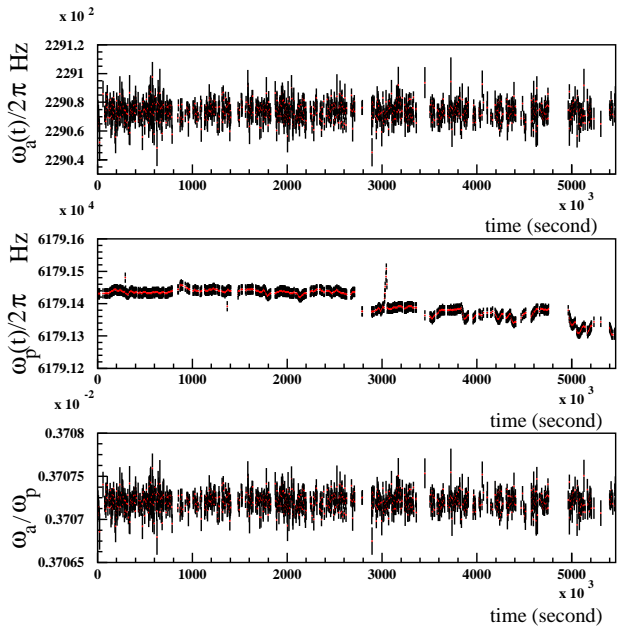}}
\caption{ 
Values of $\omega_{a}(t)$, $\omega_{p}(t)$ and
 ${\mathcal R}(t) \equiv \frac{\omega_{a}}{\omega_{p}}(t)$
from the 2001 $\mu^-$ run. The uncertainty on each $\omega_a$ 
point is about 20~ppm. }
\label{fg:omegaa01}
\end{center}
\end{figure}

The second signature of Lorentz and CPT violation would be a 
variation
of ${\mathcal R}(t)$ with a period of
one sidereal day.  The 
data are collected in `runs' of approximately 60-minute
duration, each with a time stamp at the beginning and end of run.
Data from each of these different time intervals are fitted to
the 5-parameter function (Eq.~\ref{eq:fivepar}) 
to obtain $\omega_{a}$, with the center of the time 
interval assigned as the time of that interval. The average 
magnetic field, $\omega_{p}$, for that time interval is 
used to determine ${\mathcal R}(t)$. Fig.~\ref{fg:omegaa01}
shows  $\omega_{a}(t)$, $\omega_{p}(t)$ and ${\mathcal R}(t)$ as a
function of time for the 2001 data collection period.

A sidereal variation of $\omega_a$ can be written as
$\omega_{a}(\omega_{p}(t_{i}),t_{i})=
K\omega_{p}(t_{i})+A_{\Omega}\cos(\Omega t_i +\phi)$. Dividing
by $\omega_p$ gives
\begin{equation}
{\mathcal R}(t_{i})=K+\frac{A_{\Omega}}{\omega_{p}(t_{i})}\cos(\Omega
t_i +\phi)\, ,
\label{eq:sid-freq}
\end{equation}
where $K=\lambda a_{\mu}/(1+a_{\mu})$ is a constant,
 and $A_{\Omega}$ is the amplitude of 
the sidereal variation with the sidereal period $2 \pi / \Omega$.
Two analysis techniques were used to search for an oscillation 
at the sidereal frequency: a multi-parameter fit
to Eq.~\ref{eq:sid-freq}, and the Lomb-Scargle test~\cite{Lomb-Scargle},
a spectral analysis technique developed
for unevenly sampled data such as those displayed
in Fig.~\ref{fg:omegaa01}.   With evenly sampled data
 it reduces to the usual Fourier analysis. For the time series
 $\{{h_{i}}\}$ with $ i=1, \dots\  N$,
the  normalized Lomb power at frequency $\omega$ is defined as
\begin{equation}
\begin{split}
  P_{N}(\omega) \equiv \frac{1}{2\sigma^{2}}
&
\{\frac{\left[\sum^N_{i=1}(h_{i}-\bar{h})\cos[\omega(t_{i}-\tau)]
\right]^{2}}{\sum^N_{i=1}\cos^{2}[\omega(t_{i}-\tau)]}\\
&+\frac{\left[\sum^N_{i=1}(h_{i}-\bar{h})\sin[\omega(t_{i}-\tau)]\right]^{2}}
{\sum^N_{i=1}\sin^{2}[\omega(t_{i}-\tau)]}
\}, 
\end{split}
\label{eq:lombpowerprob} 
\end{equation} 
where $\bar{h}$, $\sigma$ and $\tau$ are defined as:
\begin{equation}
\begin{split}
  \bar{h} &\equiv \frac{\sum^N_{i=1}h_{i}}{N}, \quad
\sigma^{2} \equiv \frac{1}{N-1}\sum^N_{i=1}(h_{i}-\bar{h})^{2},\\
 \tan (2\omega\tau) &\equiv 
\left( \sum^N_{i=1}\sin 2\omega t_{i}/ \sum^N_{i=1}\cos 2\omega t_{i}\right).
\end{split}
\end{equation} 
In searching for a periodic signal, the Lomb power is calculated
over a set of frequencies. For a single frequency, with no corresponding
periodic signal, the Lomb power is
distributed exponentially with unit mean. If $M$ {\it independent} frequencies
are scanned, the probability that none of them are characterized by
a Lomb power greater than $z$ is $(1-e^{-z})^M$, assuming there is no
signal present.  The significance (confidence) 
level of any peak in $P_N(\omega)$
is $1-(1-e^{-z})^M$, which is the probability
of the frequencies being scanned giving
a Lomb power greater than $z$ due to a statistical fluctuation.  
A small
value of this probability therefore indicates the presence of a significant
periodic signal. In the case of equally separated points, the number of
independent frequencies is almost equal to the number of time values.
The lowest independent frequency, $f_0$, is the
inverse of the data's time span and
the highest is roughly $(N/2)f_0$, but, because of
the uneven time sampling, may be somewhat greater. More
generally, the number of independent frequencies, which depends on the
number
and spacing of the points, as well as the number of frequencies scanned,
can be determined by Monte
Carlo simulation, using Eq.~\ref{eq:lombpowerprob} to fit for M.

\begin{figure}[]
\begin{center}
{\includegraphics[width=0.5\textwidth]{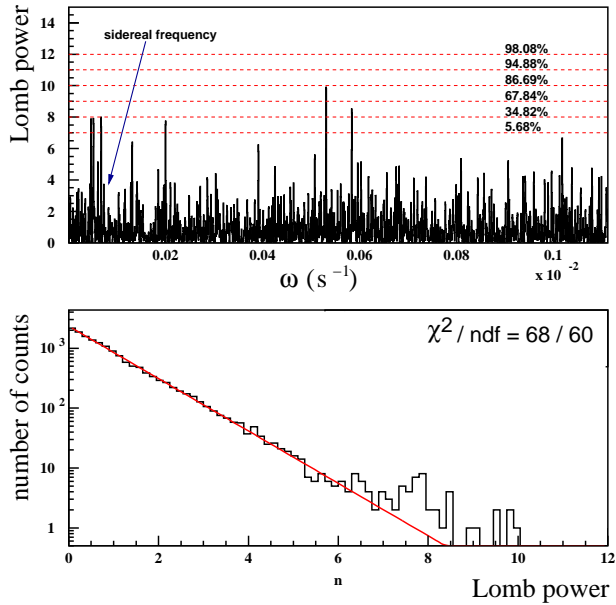}}
\caption{The Lomb-Scargle test on the 2001 data (top plot). The horizontal 
lines show the confidence level associated with
each  Lomb power.  At
 the sidereal frequency  the Lomb power is 3.4, corresponding
  confidence level less than $ 10^{-3}$\%.
 The Distribution of the Lomb power of the scanned frequencies (bottom plot)
 is consistent with an exponential distribution, indicating there is no time 
varying signal in the real data.}
\label{fg:LS01}
\end{center}
\end{figure}

The frequency spectrum for the 2001 data is shown in Fig.~\ref{fg:LS01}.
The Lomb power at the sidereal frequency is 3.4. The probability that this
is inconsistent with the null hypothesis is negligible.
 The Lomb power distribution of scanned frequencies (for 
$M=3144$) shown in the lower half of  Fig.~\ref{fg:LS01} 
is consistent with an exponential
distribution, indicating there is no significant time-varying
signal in the data.
The Lomb power spectrum reaches only a modest maximum of 9.93 at 
$f_{max} = 8.4 \times 10^{-5}$~Hz, and is not present in the 1999 or 2000
data sets.   
If there were no oscillation signal at $f_{max}$, then 14\% of the time the
measured power is expected to be that large. Moreover, only a few of the
Lomb
power spectra taken from data subsets reveal a peak at that frequency.

\begin{figure}[]
\begin{center}
{\includegraphics[width=0.5\textwidth]{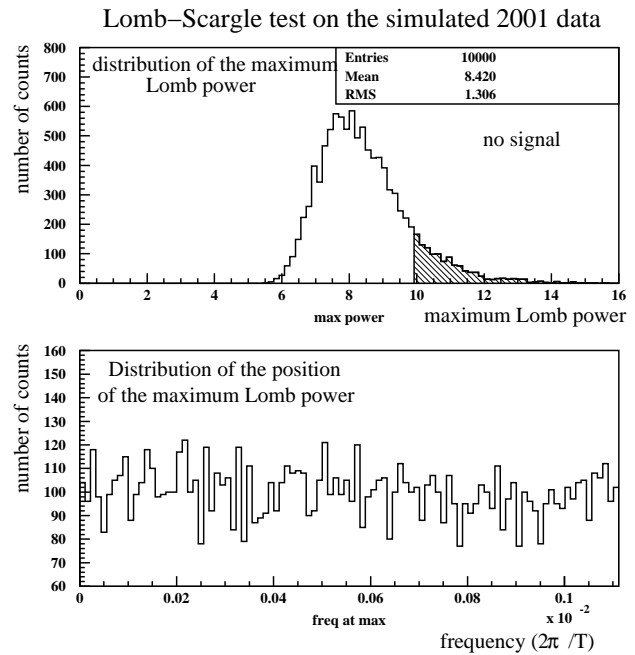}}
\caption{The Lomb-Scargle test on the simulated 2001 data with no signal. 
The shaded area shows the maximum Lomb powers greater than that of the real
 data. The locations of the maximum Lomb powers are randomly distributed
 (bottom plot).}
\label{fg:LS01sim}
\end{center}
\end{figure}

The significance of a possible sidereal signal in the
power spectrum was carefully studied
with a Monte Carlo simulation. First, a large number of artificial time
spectra
were generated. The distribution of times of the data points is chosen to
be the same as for the real data. The value of each data point was distributed
randomly with a central value equal to an average over all the actual
data values, while having 
 a standard deviation equal to that of each individual data point.
The Lomb-Scargle test was then applied to 10000 simulated data groups.
The distribution of maximum Lomb power and the corresponding frequencies are
shown in Fig.~\ref{fg:LS01sim}.

Next, sidereal oscillation signals of different amplitudes $A_\Omega$ 
were introduced
into the simulated data, and the resulting spectra were analyzed with both
the multi-parameter fit and Lomb Scargle test. Table~\ref{tb:amp} lists the
signal amplitude required so that 95\% of the simulated data sets
yield a larger $A_\Omega$ or $P(\omega)$ than the real data. The two analysis
methods give consistent results.

\begin{table}[h]
\begin{center}
\begin{tabular}{|c|c|c|}
\hline
data set & MPF amplitude & L-S amplitude\\
         & (ppm) & (ppm)\\
\hline
\hline
 1999 $\mu^{+}$ & 5.5 & 5.2\\
\hline
2000 $\mu^{+}$  & 2.2 & 2.0 \\
\hline
1999/2000 $\mu^{+}$  & 2.2 & 2.0\\
\hline
2001 $\mu^{-}$  & 4.2 & 4.2\\ 
\hline
\end{tabular}
\end{center}
\caption{The signal amplitude in ppm needed for 95\% of the simulated data
 to have larger $A_\Omega$, or $P(\Omega)$, than that of the real data.
MPF means multi-parameter fit, L-S stands for Lomb-Scargle. }
\label{tb:amp}
\end{table}

Several potential systematic effects were studied.
Since the sidereal period is very close to one
solar day, the Lomb-Scargle
test was applied to the $\omega_p$ data to check for false sidereal
variations that might be produced by diurnal temperature changes.
 The upper limits 
on the amplitude of a sidereal variation in  $\omega_{p}$ were
0.04, 0.03 and 0.08 ppm for the 1999, 2000 and 2001 data 
sets respectively, significantly smaller than the limit on 
$A_\Omega$  presented above.  Additional studies were carried out
on the 2001 data set. The data were folded back over a  four sidereal 
day time period, (i.e. modulo four sidereal days), and then analyzed.
To search for systematic effects, other sub-window time periods,
where no variation was expected, were also used,
e.g.  24 hours (solar day), or an arbitrary number 
of minutes.

No significant sidereal variation in $\mathcal R$, and hence in $\omega_a$,
is found in the E821 muon $(g-2)$ 
data.    The limits on $A_\Omega$ from the MPF in Table~\ref{tb:amp} give
at the 95\% confidence level:
\begin{equation}
\begin{split}
\check{b}_{\perp}^{\mu^{+}} &=
\sqrt{(\check{b}_{X}^{\mu^{+}})^2+(\check{b}_{Y}^{\mu^{+}})^2}
< 1.4 \times 
10^{-24}\, {\rm GeV},\\
\check{b}_{\perp}^{\mu^{-}} &=
\sqrt{(\check{b}_{X}^{\mu^{-}})^2+(\check{b}_{Y}^{\mu^{-}})^2}
< 2.6 \times 
10^{-24}\, {\rm GeV}.
\end{split}
\end{equation}
For the dimensionless figure of merit obtained by dividing by
$m_\mu$~\cite{Bluhm00}, we obtain
$r_{A_{\Omega}}^{\mu^+} \equiv 2 \sin \chi b_\perp^{\mu^+} /m_\mu < 2.0 \times 10^{-23}$ 
and  $r_{A_{\Omega}}^{\mu^-} < 3.8 \times 10^{-23}$,  which can be compared
with the ratio of the muon to Planck mass,
$m_{\mu}/M_{P}=8.7 \times 10^{-21}$.
Using the E821 and CERN~\cite{Bailey79} 
values of $\mathcal R$ for $\mu^+$ and
$\mu^-$ along with Eq.~\ref{eq:twochi}, 
we find
\begin{equation}
(m_{\mu}d_{Z0}+H_{XY}) = ( 1.8 \pm 6.0 \times 10^{-23})~{\rm GeV}.
\end{equation}
An experiment that searched for sidereal variation
 in  transitions between muonium hyperfine
 energy levels~\cite{Hughes01},  obtained $r^{\mu} \leq 5.0 \times
 10^{-22}$. Penning trap experiments with a single trapped electron 
obtained  $r^{e} \leq 1.6 \times 10^{-21}$~\cite{Mittleman99}.
  
We find no significant Lorentz and CPT violation 
signatures in the E821 muon $(g-2)$ data, which we interpret
in the framework of the standard model extension~\cite{Bluhm00}.
Limits on the parameters are of the order 
$10^{-23}$ to $10^{-24}$~GeV, with the 
dimensionless figures of merit 
$\sim 10^{-23}$.  These results represent the best test of this 
model for leptons. Both 
 $r_{\Delta\omega_{a}}^{\mu}$ and  $r_{A_{\Omega}}^{\mu}$ are much
 less than $m_\mu/M_P$, so
 E821 probes Lorentz and CPT violation signatures beyond the Planck scale.

We thank  the BNL management, along with  the staff of the 
BNL AGS for the strong support they have given the muon $(g-2)$ experiment
over a many-year period.  
We thank A. Kosteleck\'{y} for helpful comments.
This work was supported in part by the U.S. Department of Energy,
 the U.S. National Science Foundation, the German Bundesminister
 f\"{u}r Bildung und Forschung, the Alexander von Humboldt Foundation,
the Russian Ministry of Science,
 and the U.S.-Japan Agreement in High Energy Physics.

\end{document}